\documentclass[%
 reprint,
 superscriptaddress,
 amsmath,amssymb,
 aps,
 nobalancelastpage,
]{revtex4-2}
\usepackage{siunitx}
\usepackage{graphicx}% Include figure files
\usepackage{dcolumn}% Align table columns on decimal point
\usepackage{bm}% bold math
\usepackage{threeparttable,array} % manually put footnote
\newcommand{\ket}[1]{$|\mathrm{#1}\rangle$}

\begin{document}
\title{Autler-Townes Energy-Level Engineering for High-Quality All-Microwave Qubit Gate}
\preprint{APS/123-QED}

\def\SIMIT{State Key Laboratory of Materials for Integrated Circuits, Shanghai Institute of Microsystem and Information Technology, Chinese Academy of Sciences, Shanghai 200050, China}
\def\UCAS{University of Chinese Academy of Sciences, Beijing 100049, China}
\def\Shanghaitech{ShanghaiTech University, Shanghai 201210, China}
\def\ZJU{Zhejiang Key Laboratory of Micro-Nano Quantum Chips and Quantum Control, School of Physics, and State Key Laboratory for Extreme Photonics and Instrumentation, Zhejiang University, Hangzhou 310027, China}
\def\RIKEN{Center for Quantum Computing, RIKEN, Wako, Saitama 351-0198, Japan}
\def\TokyoU{Graduate School of Science, Tokyo University of Science, Shinjuku, Tokyo 162-0825, Japan}

\author{Wanpeng Gao}
\affiliation{\SIMIT}
\author{Xiaoliang He}
\affiliation{\SIMIT}
\author{Zhengqi Niu}
\affiliation{\SIMIT}
\author{Siqi Li}
\affiliation{\SIMIT}
\affiliation{\UCAS}
\author{Daqiang Bao}
\affiliation{\SIMIT}
\author{Kuang Liu}
\affiliation{\SIMIT}

% \author{\mbox{Yifan Wang}}
% \affiliation{\SIMIT}
\author{J. Q. You}
\email{jqyou@zju.edu.cn}
\affiliation{\ZJU}
\author{Jaw-Shen Tsai}
\affiliation{\RIKEN}
\affiliation{\TokyoU}
\author{Zhen Wang}
\affiliation{\SIMIT}
\affiliation{\UCAS}
% \affiliation{\Shanghaitech}
\author{Z. R. Lin}
\email{zrlin@mail.sim.ac.cn}
\affiliation{\SIMIT}
\affiliation{\UCAS}

\date{November 6, 2026}

\begin{abstract}
High-fidelity entangling gates are essential for realizing large-scale fault-tolerant superconducting quantum computing, yet existing approaches are limited: frequency-tuning techniques induce decoherence, while all-microwave-driven schemes suffer the trade-off between gate speed and leakage rate. Here, we introduce an energy-level engineering technique that employs the Autler-Townes effect to manipulate the nonlinearity of qubits without perturbing the computational subspace or degrading coherence time. This approach enables an experimentally validated, fast (97\,ns) and high-fidelity (99.40(9)\%) CZ gate, where error analysis reveals that gate errors arise mainly from static decoherence. The scheme facilitates fast, high-quality all-microwave gate operations with a simple topology and can be applied across various quantum platforms.

\end{abstract}

%\keywords{Suggested keywords}%Use showkeys class option if keyword
                              %display desired
\maketitle
High-fidelity entangling gates are the cornerstones of large-scale, fault-tolerant quantum computing \cite{DiVincenzo_2000,Krantz2019,Kwon2021}. The transmon qubit based on high-integrity superconducting circuits  \cite{Arute2019QuantumSU,Cao2023,Chao2019,Krinner_2022,Bravyi_2022,Rigetti2018} is one of the most promising platforms for constructing high-fidelity two-qubit gates due to its long coherence time and high connectivity. Superconducting two-qubit gates function by energy-level engineering to manipulate interactions involving computational states \cite{Barends_2014,PhysRevLett.125.240503,PhysRevLett.125.240502,PhysRevX.11.021058,PhysRevApplied.14.024070,PhysRevLett.127.080505,PhysRevB.81.134507,PhysRevLett.107.080502,PhysRevLett.127.200502,PhysRevA.102.062408,Sete_2021,PhysRevApplied.14.044039,Shirai_2023,PhysRevApplied.19.034071}. Yet prevailing approaches are limited: flux-tuning methods, which employ frequency modulation for entanglement generation, compromise system coherence and flux noise immunity \cite{Barends_2014,PhysRevLett.125.240503,PhysRevLett.125.240502,PhysRevX.11.021058,PhysRevApplied.14.024070,PhysRevLett.127.080505}; on the other hand, all-microwave-driven schemes, which activate indirect interactions via adjustable virtual energy states, suffer a fundamental trade-off between gate speed and leakage rate \cite{PhysRevB.81.134507,PhysRevLett.107.080502,PhysRevLett.127.200502,PhysRevA.102.062408,Sete_2021,PhysRevApplied.14.044039,Shirai_2023,PhysRevApplied.19.034071}.  

An energy shift induced by the Autler-Townes (AT) effect provides a potential solution to the aforementioned challenges. The AT effect employs microwave drive to produce selective energy-level splitting via resonance with the target transition frequency \cite{PhysRevA.81.053836,PhysRevA.89.063822,PhysRevB.88.060503}, thereby performing a drive-amplitude-dependent adjustment of the aiming state \cite{PhysRevLett.130.193603}, while the system's large anharmonicity ensures that irrelevant states remain unaffected. This technique is general and can be integrated into other quantum platforms.

In this letter, we propose a novel energy-level engineering technique based on the AT effect for building a high-quality all-microwave superconducting two-qubit gate. The two-qubit operation is activated by applying a microwave pulse close to the \ket{2} $\leftrightarrow$ \ket{3} transition frequency, adjusting the effective anharmonicity without perturbing the computational subspace or degrading coherence time due to large detuning from the qubit frequency. Leveraging the strong direct \ket{11} $\leftrightarrow$ \ket{20} interaction, we experimentally demonstrate a fast (97\,ns) and high-fidelity (99.40(9)\%) non-adiabatic CZ gate with a simple, fixed-frequency transmon topology. An optimized AT coupler scheme that can be integrated into the prevailing coupler structure is also discussed.

\begin{figure}[t]
\includegraphics[width=0.5\textwidth]{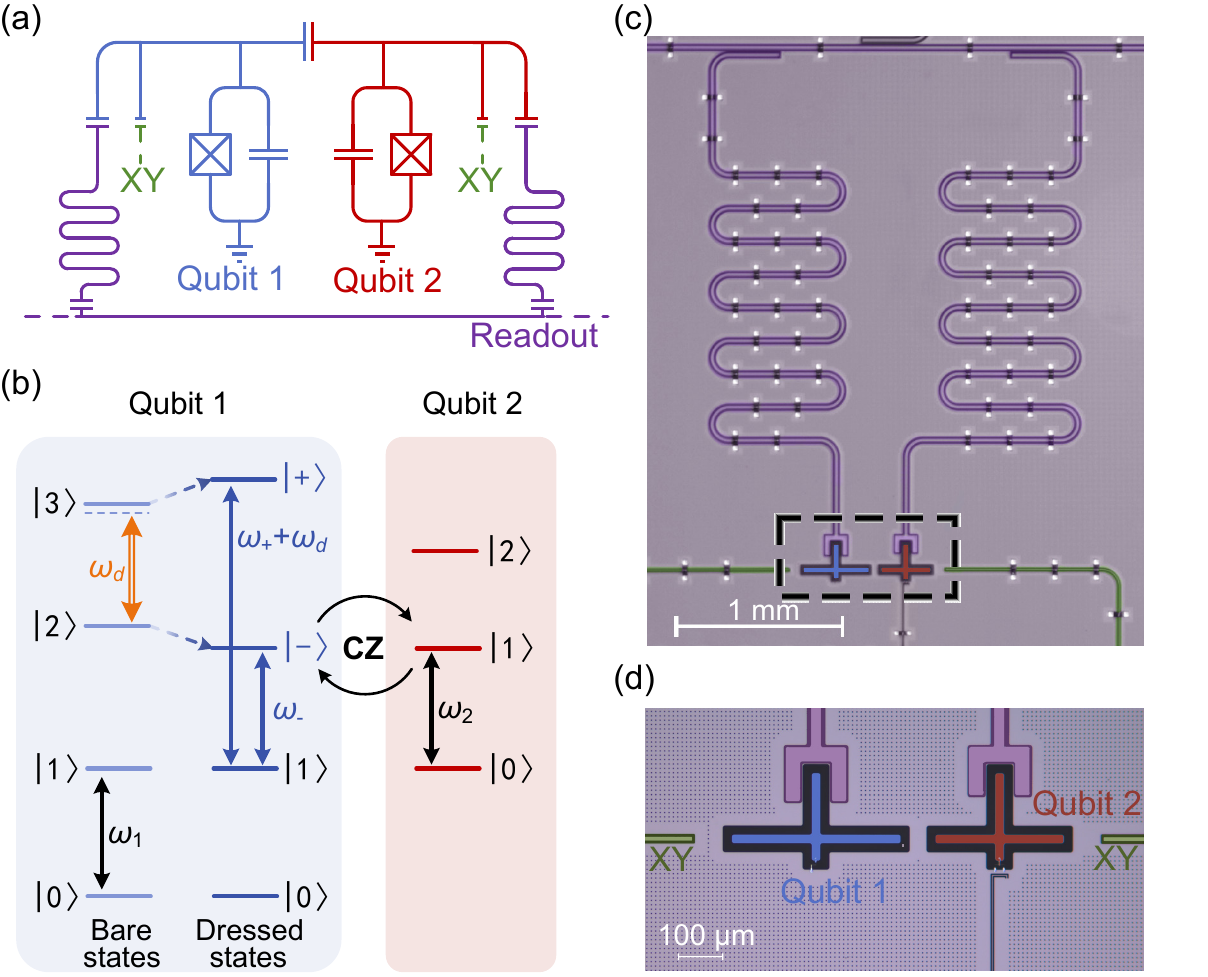}
\caption{Schematic and concept of the proposed CZ gate scheme and false-color optical image of the fabricated device. \textbf{(a)} Equivalent circuit diagram of a system consisting of a pair of coupled fixed qubits. \textbf{(b)} Energy-level diagram of the system with a microwave drive. The excited states \ket{2} and \ket{3} of $Q_1$ will split into AT doublets \ket{+} and \ket{-} under the strong microwave drive. A population exchange can take place between the \ket{-} state of $Q_1$ and the \ket{1} state of $Q_2$. \textbf{(c)} Optical image of a fabricated superconducting circuit with a pair of coupled Xmons in false colors. Most of the structures are made from Nb/Ta electrodes (light) on a sapphire substrate (dark). Other false colors correspond to circuit components displayed in \textbf{(a)}. \textbf{(d)} Detailed structure of the qubit area within the dashed box in \textbf{(c)}.}
\label{fig:Fig1}
\end{figure}

The scheme employs a pair of transversely-coupled qubits, as illustrated in Fig.~\ref{fig:Fig1}(a), with circuit Hamiltonian modeled as two coupled Duffing oscillators:
\begin{eqnarray}
\label{eq:eq1}
\hat{H}/\hbar=&&\sum\limits_{i=1,2}\left(\omega_i\hat{a}_i^\dag\hat{a}_i+\frac{\alpha_i}{2}\hat{a}_i^\dag\hat{a}_i^\dag\hat{a}_i\hat{a}_i\right)\nonumber\\
&&+g(\hat{a}_\mathrm{1}^\dag\hat{a}_\mathrm{2}+\hat{a}_\mathrm{1}^\dag\hat{a}_\mathrm{2}),
\end{eqnarray}
where  $\hat{a}_i$ and $\hat{a}_i^\dag$ are the annihilation and creation operators defined in the eigenbasis of the $i$-th Duffing oscillators, $\omega_i$ and $\alpha_i$ are the resonant frequency and anharmonicity of each oscillators, and $g$ is the transverse coupling strength between two oscillators. The system is in the dispersive regime $|g/\Delta_c|\ll1$, where $\Delta_c=\omega_\mathrm{1}-\omega_\mathrm{2}$.

To modulate the energy levels of $Q_1$, a microwave AT drive with amplitude $\Omega_\mathrm{AT}$ at frequency $\omega_d$, which is close to the \ket{2} $\leftrightarrow$ \ket{3} transition, is applied to it. The effective Hamiltonian of $Q_1$ in the frame rotating with $\omega_d$ is 
\begin{equation}
\label{eq:eq2}
\hat{H}/\hbar=\Delta|\mathrm{2}\rangle\langle\mathrm{2}|+(\Delta+\Delta_d)|\mathrm{3}\rangle\langle\mathrm{3}|+\Omega_{\mathrm{AT}}(|\mathrm{3}\rangle\langle\mathrm{2}|+|\mathrm{2}\rangle\langle\mathrm{3}|)
\end{equation}
where $\Delta=\omega_{\mathrm{12}}-\omega_d$ is the detuning between the microwave drive and the \ket{1} $\leftrightarrow$ \ket{2} transition frequency, $\Delta_d=\omega_{\mathrm{23}}-\omega_d$ is the detuning between the microwave drive and the \ket{2} $\leftrightarrow$ \ket{3} transition frequency. Due to the large detuning between the AT drive frequency $\omega_d$ and the \ket{0} $\leftrightarrow$ \ket{1} transition frequency $\omega_{\mathrm{01}}$, computational subspace is almost unaffected, thereby the corresponding states are omitted in Eq.~(\ref{eq:eq2}) for simplicity. The excited states \ket{2} and \ket{3} will become resonate in the frame rotating with $\omega_d$, and further split into AT doublets \ket{+}$=\cos\theta$\ket{2}$+\sin\theta$\ket{3} and \ket{-}$=\sin\theta$\ket{2}$-\cos\theta$\ket{3}, with $\tan\theta=\Omega_\mathrm{AT}/(\sqrt{\Delta_d^2+\Omega_\mathrm{AT}^2}-\Delta_d)$. 
% A tunable qubit is encoded in \ket{0} and \ket{-} (\ket{+} is also possible). 
The transition frequency between \ket{1} and \ket{\pm} is $\omega_{\pm}=\omega_{12}+\Delta_d/2\pm\sqrt{\Delta_d^2+\Omega_\mathrm{AT}^2}/2$. This driven system acts as a qubit with tunable anharmonicity, allowing the CZ gate to be implemented by rapidly adjusting the AT drive amplitude $\Omega_\mathrm{AT}$ to bring \ket{+} (or \ket{-}) into resonance with \ket{1} of $Q_2$, as shown in Fig.~\ref{fig:Fig1}(b). The speed of this CZ gate is determined by the effective coupling strength $\tilde{g}$ between \ket{+0} (or \ket{-0}) and \ket{11}, where the effective coupling strength satisfies $\sqrt{2}g\lesssim\tilde{g}\lesssim\sqrt{3}g$, depending on the mixing angle $\theta$ of the AT doublet state involved.

\begin{figure}[t]
\includegraphics[width=0.5\textwidth]{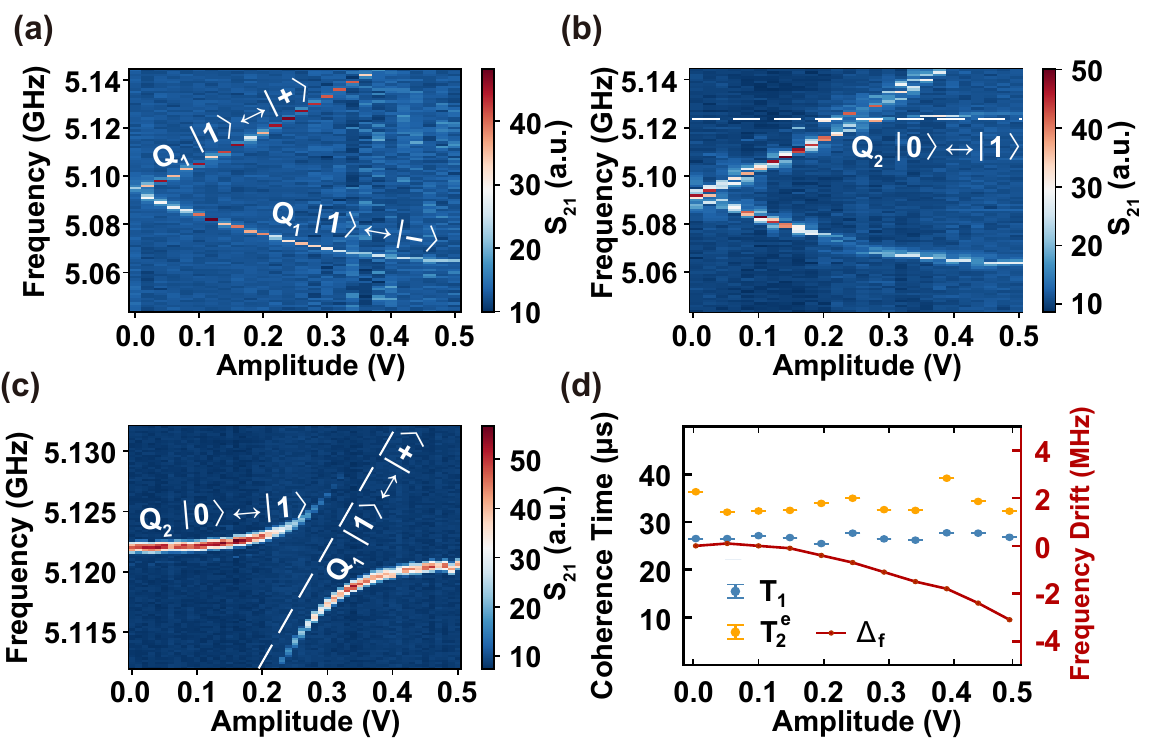}
\caption{\label{fig:fig2}Two-tone spectroscopy of the system with varying AT drive amplitude. \textbf{(a)} Two-tone spectroscopy of $Q_1$ ($f_{\mathrm{12}}$) with varying AT drive amplitude $\Omega_\mathrm{AT}$ while $Q_2$ is far detuned. \textbf{(b)}  Two-tone spectroscopy of $Q_1$ ($f_{\mathrm{12}}$) with varying AT drive amplitude $\Omega_\mathrm{AT}$ while $Q_2$ is fixed at working point, the avoided crossing indicates the coupling between the \ket{11} state and the \ket{+0} state. \textbf{(c)} Detailed spectrum of the avoided crossing area acquired by measuring $Q_2$ spetroscopy. \textbf{(d)} The impact of AT drive on $Q_1$ by measuring $Q_1$ frequency (red line) and coherence times $T_1$ (blue dots) and $T_{2}^e$ (yellow dots) with varying AT drive amplitude $\Omega_\mathrm{AT}$. }
\end{figure}

A superconducting quantum chip consisting of two coupled Xmon qubits has been fabricated for experimental validation \cite{Barends2013}, as shown in Figs.~\ref{fig:Fig1}(c) and (d). Xmon qubits are modified transmons featuring a cross-type electrode that provides large shunt capacitance to improve both scalability and charge noise immunity \cite{Koch2007,You2007}. In this experiment, $Q_2$ is biased at a fixed point and therefore can be treated as a fixed frequency qubit, though it was fabricated to be tunable for other purposes. The device parameters are characterized as follows: Xmons frequencies are $\omega_1/2\pi\simeq$ 5.272\,GHz and $\omega_2/2\pi\simeq$ 5.124\,GHz; Xmons anharmonicities are $\alpha_1/2\pi\simeq$ -178\,MHz and $\alpha_2/2\pi\simeq$ -205\,MHz. The measured energy relaxation times $T_1$ are $27\,\SI{}{\micro\second}$ and $17\,\SI{}{\micro\second}$, and the echo dephasing times $T_2^e$ are $36\,\SI{}{\micro\second}$ and $22\,\SI{}{\micro\second}$ for the two qubits, respectively.

\begin{figure}[t]
\includegraphics[width=0.5\textwidth]{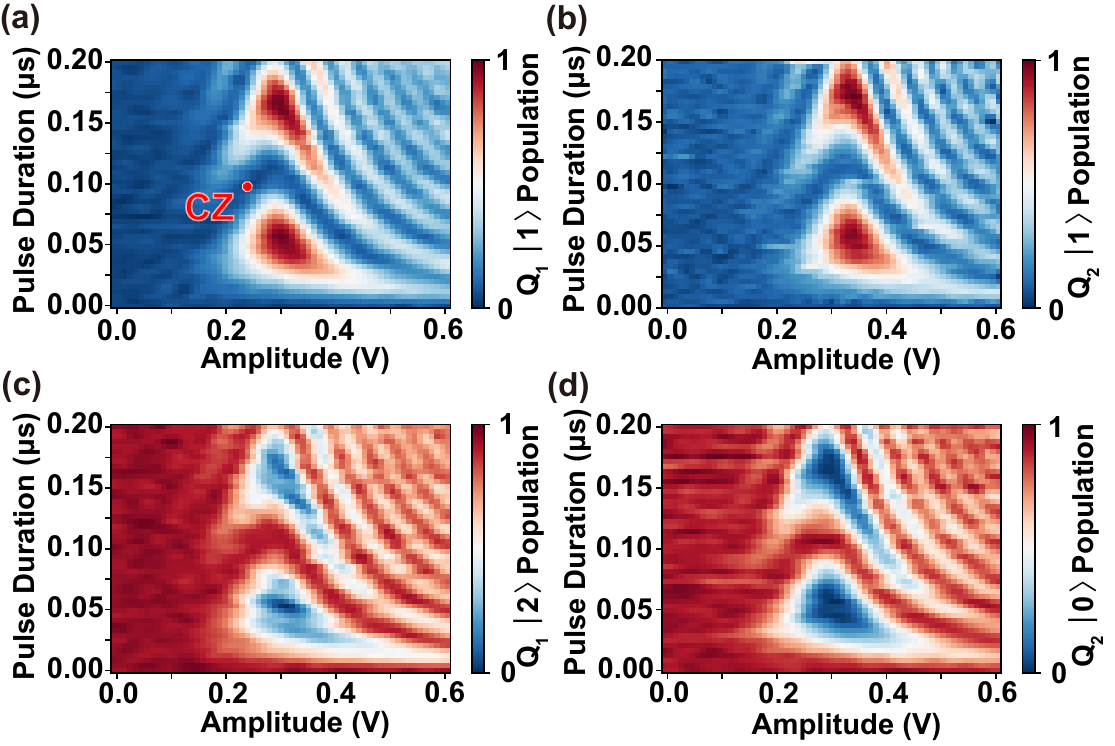}
\caption{\label{fig:fig3} CZ gate calibration. The system starts from the \ket{11} state, and the population exchange between the \ket{11} state and the \ket{20} state with varying AT drive amplitude $\Omega_\mathrm{AT}$ and pulse length $\tau$ is monitored. The experiment results of population remained in the \ket{11} state are displayed in \textbf{(a,b)}, and the results of population transferred to the \ket{20} state are displayed in \textbf{(c,d)}. The red point in \textbf{(a)} corresponds to the CZ operation point. The underlying data is interpolated onto a finer grid, with minor subsequent smoothing for presentation.}
\end{figure}

The energy-level engineering technique is first demonstrated by measuring the dependence of qubit two-tone spectroscopy on the AT drive amplitude $\Omega_\mathrm{AT}$. The system is prepared in the \ket{10} state, and the AT drive frequency $\omega_d/2\pi$ is set to 4.904\,GHz, which is near resonance with the \ket{2} $\leftrightarrow$ \ket{3} transition frequency of $Q_1$. Figure.~\ref {fig:fig2}(a) shows the two-tone spectroscopy results while $Q_2$ is far detuned. The frequency difference between the AT doublets of $Q_1$ increases with the AT drive amplitude $\Omega_\mathrm{AT}$, proving the tunability of the qubit anharmonicity. While $Q_2$ is fixed at 5.124\,GHz, the avoided crossing between the upper branch and $Q_2$ frequency indicates the transverse coupling between the qubits [Fig.~\ref{fig:fig2}(b)]. In terms of computational subspace, the \ket{10} $\leftrightarrow$ \ket{11} transition frequency is pushed away at one point when adjusting the AT drive amplitude $\Omega_\mathrm{AT}$ [Fig.~\ref{fig:fig2}(c)], and the effective coupling strength measured is $g_\mathrm{mea}/2\pi\simeq5.9$\,MHz, which supports the realization of a fast CZ gate. Figure.~\ref{fig:fig2}(d) shows the impact of AT drive on the computational subspace of $Q_1$. The coherence times $T_1$ and $T_{2}^e$ of $Q_1$ remain almost unaffected by the varying AT drive amplitude. Though, a frequency drift ($\lesssim$ 2\,MHz around the CZ working point shown in Fig.~\ref{fig:fig3}) which may arise in the off-resonant drive coupled to $Q_1$'s \ket{1} $\leftrightarrow$ \ket{2} transition is observed, the parasitic phase accumulation assisted with this frequency drift can be compensated with a virtual-Z operation on $Q_1$ after the CZ gate operation. A virtual-Z operation is also applied on $Q_2$ after the CZ gate to compensate the single-qubit Z errors arising from the frequency tuning process and the off-resonant coupling between the \ket{1} $\leftrightarrow$ \ket{2} transition frequency of $Q_2$ and the AT drive field through on-chip microwave crosstalk.

A CZ gate calibration is then performed within the avoided-crossing section shown in Fig.~\ref{fig:fig2}(c). The system starts from the \ket{11} state by applying $\pi$ pulse to both qubits. Then, the AT pulse amplitude $\Omega_\mathrm{AT}$ is rapidly increased to bring the \ket{11} state and the \ket{+0} state into resonance. After waiting for a varying delay time $\tau$, the AT pulse amplitude $\Omega_\mathrm{AT}$ is quickly decreased to zero. After this process, the system evolution is tracked by measuring populations of $Q_1$'s \ket{1} state [Fig.~\ref{fig:fig3}(a)], $Q_2$'s \ket{1} state [Fig.~\ref{fig:fig3}(b)], $Q_1$'s \ket{2} state [Fig.~\ref{fig:fig3}(c)], and $Q_2$'s \ket{0} state [Fig.~\ref{fig:fig3}(d)] versus $\Omega_\mathrm{AT}$ and $\tau$. These Chevron patterns indicate a coherent \ket{11} $\leftrightarrow$ \ket{+0} exchange during the process, and the system acquires a geometric phase after a full cyclic evolution. The gate parameters are marked with a red dot in Fig.~\ref{fig:fig3}(a), and a moderate deviation from the middle is required for the residual ZZ interaction compensation during the gate. A fast 97\,ns CZ gate is achieved, showcasing the direct interaction advantage of the scheme.

\begin{figure}[t]
\includegraphics[width=0.5\textwidth]{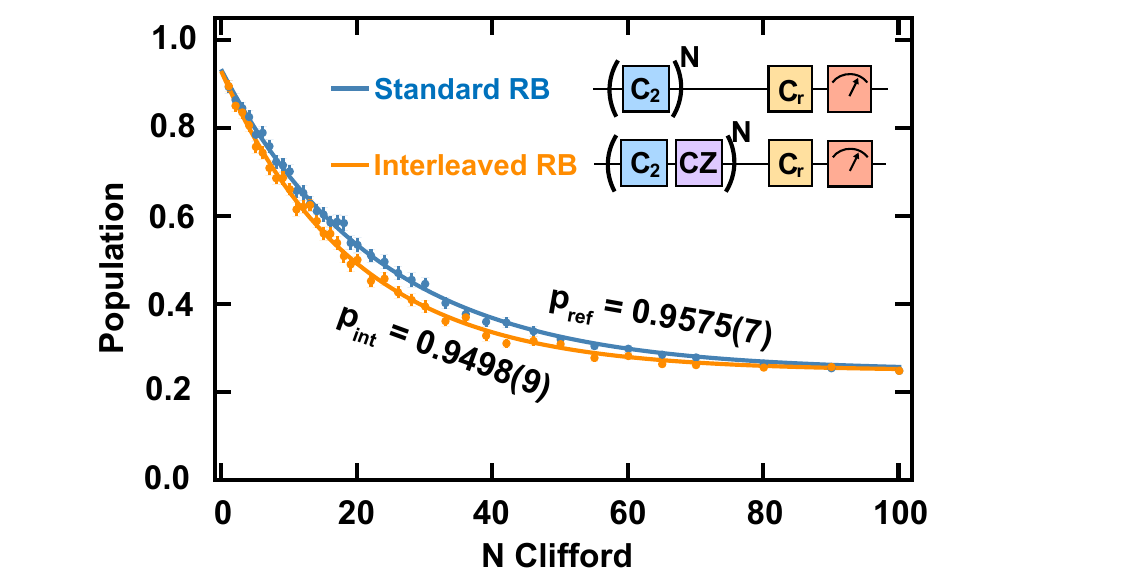}
\caption{\label{fig:fig4} Reference and interleaved randomized benchmarking (RB) results of the CZ gate. Blue and orange dots are the averaged experimental results of the reference RB and interleaved RB, respectively. 300 random sequences for each sequence length are measured, and the error bars represent 95\% confidence. The fitted fidelity per Clifford is $p_\mathrm{ref}=0.9575(7)$ for reference sequence, $p_\mathrm{int}=0.9498(9)$ for interleaved sequence, and extracted CZ gate fidelity is $F_\mathrm{CZ}=0.9940(9)$.}
\end{figure}

Fidelity assessment of this CZ gate is performed by Clifford-based Randomized Benchmarking (RB) \cite{PhysRevA.87.030301,Barends_2014}. A reference RB experiment is performed by applying multiple random sequences composed of two-qubit Clifford gates $(C_2)$ followed by a recovery gate $(C_r)$ to invert the overall operation, and the results are averaged over 300 random sequences. An interleaved RB experiment is conducted in the same way, with additional CZ gates inserted after each reference RB sequence. Figure.~\ref{fig:fig4} shows the measured sequence fidelity versus the number of two-qubit Cliffords $m$ for both reference and interleaved cases. Using an exponential fit, the fidelity per Clifford gate in the reference and interleaved case is $p_\mathrm{ref}=0.9575(7)$ and $p_\mathrm{int}=0.9498(9)$, respectively. By comparing these two traces, the CZ-gate error rate is $r_\mathrm{CZ}=\frac{3}{4}\left(1-\frac{p_\mathrm{int}}{p_\mathrm{ref}}\right)=0.0060(9)$ and fidelity is $F_\mathrm{CZ}=1-r_{\mathrm{CZ}}=0.9940(9)$, which surpasses the surface code error correction threshold \cite{Fowler2012}.
To further improve fidelity, the CZ gate error budget is estimated. The decoherence error is estimated by analyzing the energy relaxation time $T_1$ and pure dephasing time $T_\phi$ while qubits remain in the idle state. The decoherence errors induced by the relaxation and pure dephasing of each qubit are modeled as $r_\mathrm{relaxation}=\frac{2}{5}\tau_\mathrm{gate}/T_1$ and $r_\mathrm{dephasing}=\frac{2}{5}\tau_\mathrm{gate}/T_\phi$ \cite{PhysRevX.14.041050}, respectively. The total contribution of the decoherence error is $0.0047$. Thus, the static decoherence error accounts for approximately $\sim78\%$ of the total gate error, indicating that gate performance can benefit from an enhanced coherence time. To estimate the control error, we also numerically calculated the effect of the AT driving field on the computational subspace of $Q_1$. A simple calculation on an off-resonant driven qutrit yields an average parasitic X-rotation angle of 0.08 rad, which corresponds to an gate infidelity of 0.0013. This takes up most of the remaining $\sim22\%$ error, that could be neutralized by an optimized pulse shape.

\begin{figure}[t]
\includegraphics[width=0.5\textwidth]{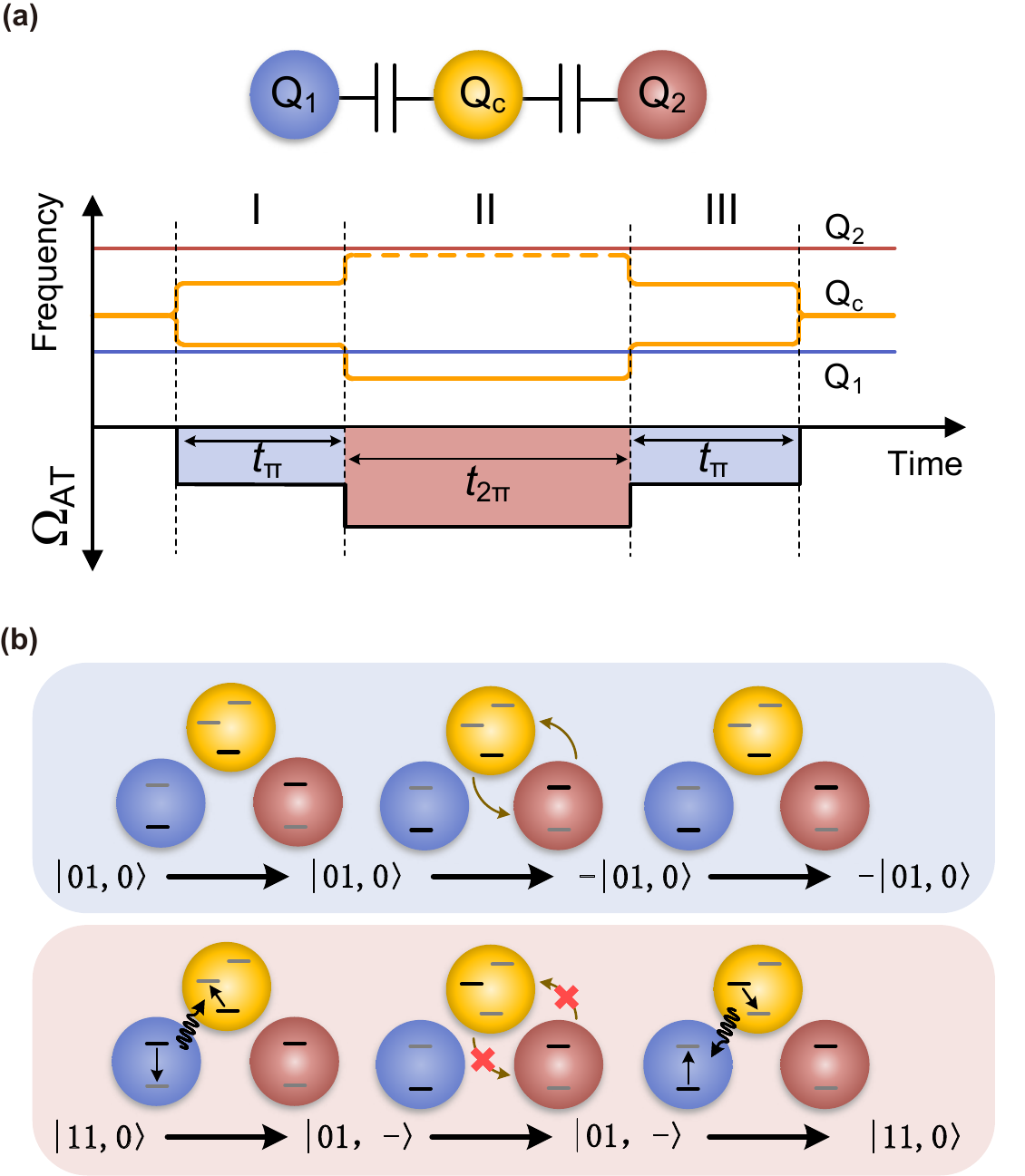}
\caption{\label{fig:fig5} Schematic of the AT coupler scheme. \textbf{(a)} The scheme comprises two data qubits $Q_1$, $Q_2$ and an AT coupler $Q_c$ that cancels residual coupling between data qubits (above). The pulse sequence of the AT drive in this scheme consists of three steps (middle). The eigenenergies of $Q_1$, $Q_2$ and $Q_c$ are plotted in blue, red and yellow lines, respectively. The amplitude of AT drive (beneath) is set to bring $Q_1$ and the lower branch of $Q_c$ into resonance during step I and III, and bring $Q_2$ and the upper branch of $Q_c$ into resonance in step II. \textbf{(b)} Schematic representation of the entangling operation. The coupler's \ket{+} state interacts with $Q_2$, resulting in a controlled phase during step II, only when $Q_1$ starts from the \ket{0} state. The coupler collects the population of $Q_1$ in step I and returns it in step III regardless of the state of $Q_2$. This process is equivalent to a CZ gate with single-qubit Z rotations. }
\end{figure}

This AT energy-level engineering technique is also compatible with the commonly used coupler structure for large-scale quantum processors \cite{PhysRevApplied.10.054062}. A coupler is added to cancel residual coupling between data qubits as shown in Fig.~\ref{fig:fig5}(a), and the coupler also has a V-shaped energy structure of AT doublets. The gate is implemented in three steps [Fig.~\ref{fig:fig5}(b,c)]: (I) A square-shaped AT drive is applied to the coupler and maintained until the activated population exchange between $Q_1$ and the lower branch is complete; (II) The amplitude of AT drive is increased to generate an intersection between $Q_2$ and the upper branch, allowing $Q_2$ to accumulate conditional phase during the process; (III) The AT drive amplitude is reduced to the value in step I, and persists for another population exchange between $Q_1$ and the lower branch. The concept of this gate is illustrated in Fig.~\ref{fig:fig5}(c): $Q_2$ may acquire the phase $\pi$ in step II if $Q_1$ starts from \ket{0}, or remain unchanged if $Q_1$ starts from \ket{1} and occupies the coupler between step I and step III, while $Q_1$ always returns to the initial state with excess phase regardless of the state of $Q_2$, which is equivalent to a CZ gate with single-qubit Z rotations. Since the AT coupler acts as a switch with $Q_1$ controlling the existence of the upper branch, the tunable upper branch could also be utilized to generate entanglement between $Q_2$ and a third qubit $Q_3$, and the whole process is equivalent to a Toffoli gate with single-qubit Z rotations. This approach uses fewer two-qubit gates than the standard Toffoli gate, thereby optimizing hardware mapping for quantum algorithms.

In conclusion, an energy-level engineering technique based on the AT effect is proposed and experimentally validated in a superconducting quantum circuit. This approach can adjust the nonlinearity of the qubit while protecting the computational subspace and coherence time, which allows a fast (97\,ns) and high fidelity (99.40(9)\%) all-microwave CZ gate with a simple, fixed-frequency topology. The gate fidelity is assessed via randomized benchmarking, and the gate errors are mainly attributed to the static decoherence error. Its performance can be improved by enhancing the coherence time and optimizing the pulse shape. An optimized AT coupler scheme that is compatible with the commonly used coupler structure of quantum processors is also discussed. The technique is applicable to other quantum platforms struggling with energy-level engineering and holds promise to propel the progress of large-scale, fault-tolerant quantum computing.

\begin{acknowledgments}
The authors acknowledge the use of the Superconducting Electronics Facility (SELF) at Shanghai Institute of Microsystem and Information Technology. The authors would like to thank Yasunobu Nakamura and Fei Yan for insightful and inspiring discussions. This work is supported by the National Natural Science Foundation of China (Grants Nos. 92576207, No. 92265202 and No. U25A20199), the National Key Research and Development Program of China (Grant No. 2023YFB4404904), the Key-Area Research and Development Program of Guangdong Province, China (No. 2020B0303030002).
\end{acknowledgments}

% \nocite{*}
\bibliography{AT}

\begin{appendix}
\section{Autler-Townes effect}

The Hamiltonian of a transmon qubit driven by a weak probe field of Rabi frequency $\Omega_p$ coupling the \ket{0} $\leftrightarrow$ \ket{1} transition and a strong pump field of Rabi frequency $\Omega_c$ coupling the \ket{1} $\leftrightarrow$ \ket{2} transition can be expressed as:
\begin{align}
\centering
\label{suppeq:H}
H&=  {
    \begin{pmatrix}
    0 & \Omega_pe^{-i\omega_pt} & \\
    \Omega_pe^{i\omega_pt} & \omega & \Omega_ce^{-i\omega_ct}\\
     & \Omega_ce^{i\omega_ct} &2\omega+\alpha
    \end{pmatrix}%
  }     
\end{align}
where $\omega_p$ and $\omega_c$ are the frequencies of microwave fields, $\omega$ and $\alpha$ are the frequency and anharmonicity of the qubit, respectively. In the limit of $\Omega_c/\Omega_p\gg1$, the Hamiltonian can be expressed in a frame rotating with pump field frequency $\omega_c$:
\begin{align}
\centering
\label{suppeq:H_eff}
H_{\mathrm{eff}}&=  {
    \begin{pmatrix}
    0 & \hspace*{0.8cm} & \hspace*{0.8cm}\\
     \hspace*{0.8cm}& \Delta & \Omega_c\\
     & \Omega_c &\Delta+\Delta_d
    \end{pmatrix}%
  }     
\end{align}
where $\Delta=\omega-\omega_c$ and $\Delta_d=\omega+\alpha-\omega_c$. This Hamiltonian can be diagonalized and obtain Autler-Townes Doublets \ket{+}$=\cos\theta$\ket{1}$+\sin\theta$\ket{2} and \ket{-}$=\sin\theta$\ket{1}$-\cos\theta$\ket{2} with eigenenergies $\omega_{\pm}=\omega+\Delta_d/2\pm\sqrt{\Delta_d^2+\Omega_c^2}/2$, where $\tan\theta=\Omega_c/(\sqrt{\Delta_d^2+\Omega_c^2}-\Delta_d)$. In Fig.~2 of the main text, the weak probe field of Rabi frequency $\Omega_p$ couples the \ket{1} $\leftrightarrow$ \ket{2} transition and the strong pump field of Rabi frequency $\Omega_c$ couples the \ket{2} $\leftrightarrow$ \ket{3} transition, leaving the \ket{0} $\leftrightarrow$ \ket{1} transition almost unaffected by the AT drive.

\section{Device information}
The transmon qubits and the resonators are fabricated with 100\,nm thick Tantalum/Niobium bilayer on a Sapphire (111) substrate. The substrate is first etched with vapor Hydrogen Fluorine (VHF) to remove native oxide, and the Tantalum/Niobium film is deposited with magnetron sputtering  soon afterwards. Tantalum/Niobium structures are defined with the laser direct writing lithography and etched with the reactive-ion etching process. The sample is further disposed with BOE to remove oxide from the metal surface. Josephson junctions are realized by the cross-junction defined by ZEP/PMGI photoresist bilayer with electron-beam lithography. Aluminum air-bridges are applied over each coplanar waveguide to suppress the slot line mode. The device parameters are summarized in Table~\ref{tab:tab1}. 
\begin{table}[!htbp]
%\begin{table}[htbp!] positioning is bad
%\centering % used for centering table
\caption{\label{tab:tab1}Device parameters.} % title of Tabletrans
\begin{ruledtabular}
\begin{threeparttable}
\begin{tabular}{c | c c c } % centered columns (2 columns)
%\hline \hline % inserts single horizontal line
\hspace*{3cm} & QB1 & QB2 \\\hline  \\[-0.3cm]% inserting body of the table
$\omega/2\pi$\tnote{a}~ (GHz) & 5.272 & 5.124 & \\ 
$\alpha/2\pi$\tnote{b}~ (MHz)&  -178 & -205 \\ %\hline
$g_{\mathrm{12}}/2\pi$\tnote{c}~ (MHz) & \multicolumn{2}{c}{5.9} \\ %\hline
$T_1$\tnote{d}~ (\SI{}{\micro\second}) & 27 & 17  \\ %\hline
$T_2^*$\tnote{d}~ (\SI{}{\micro\second}) & 20 & 6   \\ %\hline
$T_2^{\mathrm{echo}}$\tnote{d}~ (\SI{}{\micro\second}) & 36 & 22\\ %\hline
 $\omega_{\mathrm{r}}/2\pi$\tnote{e}~ (GHz) & 6.873 & 6.681\\ %\hline
 $\kappa_{\mathrm{r}}/2\pi$\tnote{f}~ (MHz) & 0.5 & 0.5\\ %\hline
 $\chi_{\mathrm{r}}^{(0,1)}/2\pi$\tnote{g}~ (MHz) & 0.68 & 0.65 \\ %\hline

\end{tabular}
\footnotesize
\begin{tablenotes}
\item[a] $|0\rangle \rightarrow |1\rangle$ transition frequencies at the idling configuration.
\item[b] Anharmonicities at the idling configuration.
\item[c] Pairwise coupling strengths at the idling configuration.
\item[d] Energy decay time ($T_1$), Ramsey decay time ($T_2^*$), and spin-echo decay time ($T_2^{\mathrm{echo}}$) measured at the idling configuration.
\item[e] Readout resonator frequency.
\item[f] Readout resonator linewidth.
\item[g] Effective dispersive shifts for the $|0\rangle \rightarrow |1\rangle$ transition due to the interaction with the readout cavity mode.
\end{tablenotes}

\label{supptable:device_params}
\end{threeparttable}
\end{ruledtabular}
\end{table}

\section{Measurement setup}
\begin{figure}[!htbp]
\includegraphics[width=0.5\textwidth]{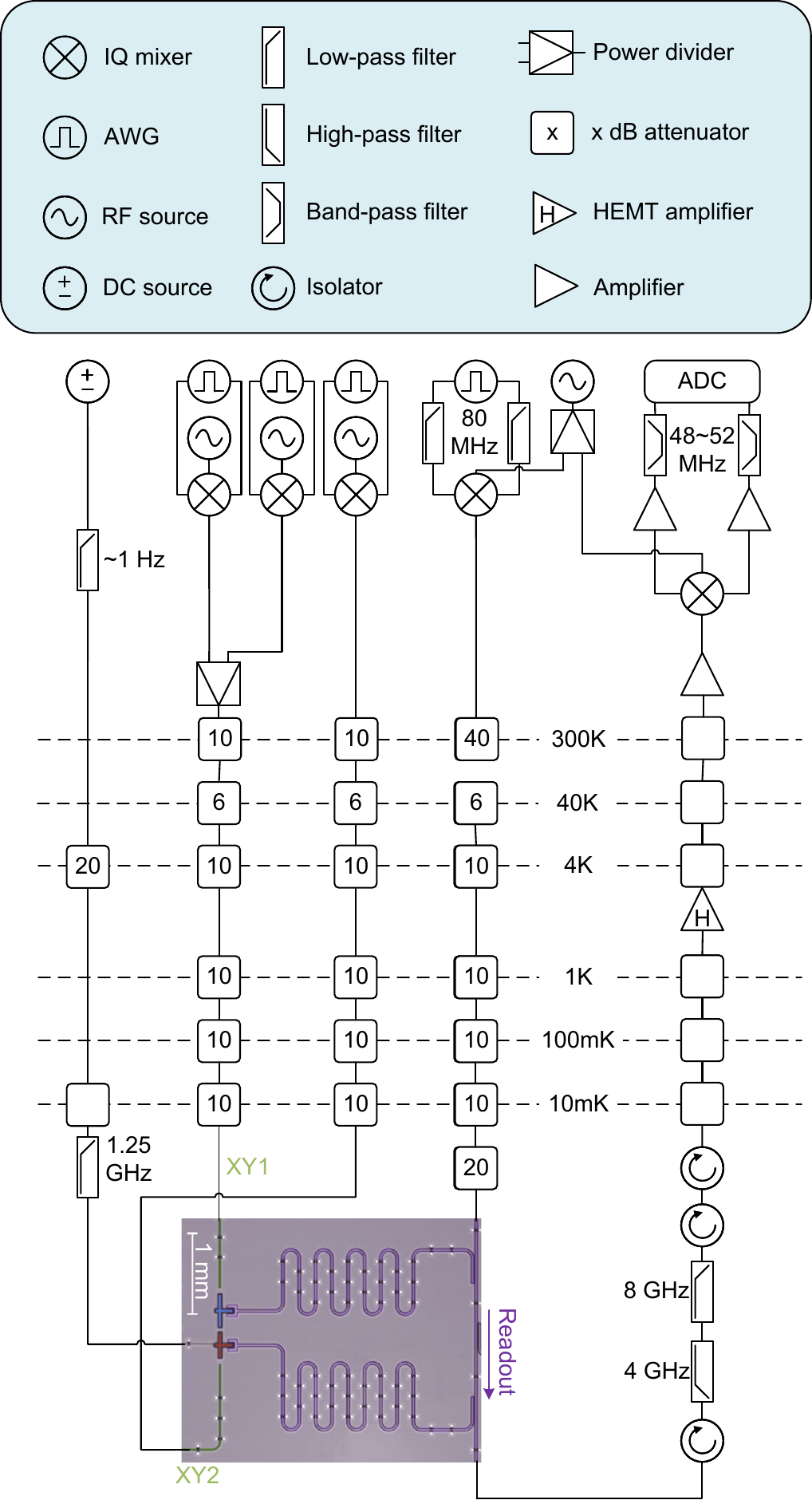}
\caption{\label{fig:figa1} A schematic diagram of the experimental setup.}
\end{figure}

The experiment is performed in a BlueFors XLD-1000 dilution refrigerator with a base temperature of about 10\,mK. As shown in Fig.~\ref{fig:figa1}, each microwave input line of the dilution refrigerator has about 52\,dB attenuation at 6\,GHz including wiring loss. The flux control line has a 20\,dB attenuator at the 4\,K stage and a low-pass filter at the 10\,mK stage. An additional RC low-pass filter is applied at the room-temperature region to suppress low-frequency current noise. Microwave pulses are generated by an arbitrary waveform generator and an IQ Mixer. The output signal is amplified with a low-noise HEMT amplifier at 4\,K stage and further amplified with a MITEQ amplifier outside the dilution refrigerator before demodulation and data processing. 

\end{appendix}

% The \nocite command causes all entries in a bibliography to be printed out
% whether or not they are actually referenced in the text. This is appropriate
% for the sample file to show the different styles of references, but authors
% most likely will not want to use it.

% \bibliography{apssamp}% Produces the bibliography via BibTeX.

\end{document}